\documentclass[12pt]{article}
\usepackage{amsmath,amssymb,amsfonts,accents,cite}
\usepackage{graphicx}
\usepackage{stmaryrd}  
\usepackage{enumerate} 

\def\nn{\nonumber}

\newtheorem{lemma}{Lemma}%[section]

\newtheorem{thm}{Theorem}

\begin{document}

\begin{center}
{\large\bf On Casimir Operators of Conformal Galilei Algebras 
}\\
~~\\

{\large Fahad Alshammari$^{1,2}$,  Phillip S. Isaac$^1$ and Ian Marquette$^1$}\\
~~\\
$^1$ 
School of Mathematics and Physics, The University of Queensland, St Lucia QLD 4072, Australia \\
$^2$ 
Department of Mathematics, Prince Sattam Bin Abdulaziz University, Saudi Arabia. \\
~~\\
{\small Email: fahad.alshammari@uqconnect.edu.au, psi@maths.uq.edu.au, i.marquette@uq.edu.au}
\end{center}

\begin{abstract}

In previous work, we introduced an algorithm that utilises differential
	operator realisations to find polynomial Casimir operators of Lie
	algebras. In this article we build on this work by applying the
	algorithm to several classes of finite dimensional conformal Galilei
	algebras with central extension. In these cases we highlight the
	utility of an algebra anti-automorphism, and give relevant details
	through key examples.

\end{abstract}

\section{Introduction}
The role of Casimir operators of Lie algebras is undeniably important in
certain areas of physics. 
For example, based on Lie symmetries of certain quantum Hamiltonians in
fundamental cases, the energy
spectrum can be expressed in terms of the eigenvalues of the Casimir elements
on an irreducible submodule of the Hilbert space. 
Explicit expressions of Casimir operators of
semisimple Lie algebras are well known. 
Such formulae for the Casimir
operators of semisimple Lie algebras have been obtained by many authors 
with some key developments being given in
\cite{ Gr64, Ok77, Po66 ,Ra49}. 
However, for non-semisimple Lie algebras, they remain largely unexplored due
to the complex nature of the algebraic structure itself.

Conformal Galilei groups and their Lie algebras are a class of nonrelativistic
algebra \cite{Ne97}. Physical applications associated with these
Galilei algebras include both classical and quantum mechanics \cite{Alv07 },
nonrelativistic spacetime and gravity \cite{Bag10,Du09,Ho10} , fluid dynamics
\cite{Zh10}, nonrelativistic holography and electrodynamics  \cite{Ai13, Le73,
Lev71,NeDR97}. Conformal Galilei algebras form a class of non-semisimple Lie
algebras, with each member characterised by two parameters $d$ (positive
integer) and $\ell$ (positive integer or positive
half-integer). The corresponding algebra is referred to as the conformal
Galilei algebra with rational dynamical exponent \cite{Du09,Du11,Ne97,St13}.
The smallest instance of $\ell=\frac{1}{2}$ is known in the literature
\cite{Ni72} as the Schr\"{o}dinger algebra. The conformal Galilei algebra with
higher value of $\ell$ is studied from physical and mathematical points of view
\cite{Ai12, Ai11, Ai13, Ai10, An12, Ba68, Ca09, Du09, Do97,  Mar10,
Ne97,Ni72, Pe77}. In many cases it has been observed that physical systems having a
connection with $\ell$-conformal Galilei ($\ell>\frac{1}{2}$) algebra are
described by Lagrangians or Hamiltonians with higher order derivatives. Also,
there is a connection with this algebra as a symmetry algebra for generalised
oscillator systems such as the Pais-Uhlenbeck oscillator \cite{ An14}.

The purpose of this paper is two fold: firstly, we use an efficient and
functional algorithm, as presented in \cite{Als17} and code using symbolic
software to show how a structural
map, namely an algebra anti-automorphism, along with a triangular decomposition
of the Lie algebra, can be combined to simplify the algorithm.
Secondly, we apply the algorithm to the conformal Galilei algebra for which, in
some cases, explicit expressions for the Casimir operators have not been
determined previously. We present certain
explicit formulas for the case of dimension $d=1,2$ and arbitrary  $\ell$. 

The contents of this paper are organized as follows. In the next section we
give a short review of generators of the $\ell$-conformal Galilei algebra and its
central extensions. In section 3 and 4 we apply our  direct approach
using differential operator realisations to the conformal Galilei algebra with central extension for
the case of $d = 1$ and half-odd integer $\ell$, and the case of $d = 2$ and arbitrary
values of $\ell$. We highlight the advantage of the structural map in further
reducing the amount of computation. 
 
\section{The conformal Galilei algebra}
The class of Lie algebras known as the  Conformal Galilei algebras have two
types of central extension according to the values of $d$ and $\ell$.
In the literature, it is  not surprising that there are many results on
central extensions of various classes of Lie algebras. Several authors have
studied  the  existence of central extensions  in the context of conformal
Galilei algebras\cite{Ai12, Ai13,Lev71,Bar54, Ho14 ,Mar10,Luk97, Luk06,
Luk7}.

Various results have already appeared in the literature concerning explicit
expressions for Casimir
operators for the cases  $d=1,2,3$ and $\ell= \frac{1}{2}$ (e.g. see 
\cite{Ai10, Ca05 , Hen97 , Pe77} ). With the help of the orbit method,
Casimir operators of  conformal Galilei algebras with mass extension have been
found in \cite{An12} for the case of $d=3$ and  half-odd integer $\ell$. For
other values of $d\geq1$ and half-odd integer $\ell\geq \frac{3}{2}$, even in
the case $d=1,2$, the Casimir operators have not been given in explicit detail.
The current article has the modest goal of filling this gap for $d=1,2$, and
various values of $\ell$ which allow a central extension. In doing so,
we discover certain structural aspects of the conformal Galilei algebras which can
improve the functionality of the algorithm presented in \cite{Als17}.

More general cases corresponding to $d\geq2$ and half-odd integer and
integer values $\ell\geq \frac{3}{2}$ will be the subject of future work. 

Let us introduce the finite-dimensional $\ell$-conformal Galilei algebra
without central extension in $d$-dimensional space, denoted $
\mathfrak{{g}}_\ell(d)$, is a semi-direct sum of
\begin{eqnarray*}
\mathfrak{sl}(2)\oplus \mathfrak{so}(d) \niplus  \mathfrak{R}^{(2\ell+1)d},     
\end{eqnarray*}
where $\mathfrak{R}^{(2\ell+1)d}$  is the abelian ideal relating to a space-translations, Galilean boosts and higher order accelerations. 
 This algebra is parametrized by  $d$ and $\ell$ which take on values
\begin{eqnarray*}
d = 1, 2, 3, . . ., \qquad \ell=\frac{1}{2}, 1, \frac{3}{2}, 2, \frac{5}{2}, . . .
\end{eqnarray*}
%The dimension is given by $2n\ell +4+\frac{n(n+1)}{2}$.
Generally, the $\mathfrak{{g}}_\ell(d)$ is specified by the following generators( see\cite{Ai12,Ai11,Ai13,Ai10}).
\begin{equation*}
	{ P_{n,i}, H, D, C, J_{ij}=-J_{ji}; \,\ n = 0,1, 2, \ldots, 2\ell; \,\
	i,j =1, 2, \ldots,  d.}
\end{equation*}
The table below shows these generators can be classified as a set of transformations in the context of Galilean relativity (i.e. before the introduction of Einstein's special relativity)  \cite{ Mar10,Ne97}.

\begin{table}[htbp]
\centering
\caption{Coordinate transformations  in Galilean relativity}
{\small
    \begin{tabular}{| p{5.4cm} |p{2.68cm} | p{1.75cm}|p{4.45cm} |p{1cm} |}
 \hline
 Coordinate transformations& Linearisation & Generator &Physical description \\
 \hline 
$(x_i \to x_i + \epsilon x_j , x_j \to x_j - \epsilon x_i )$ & $-x_i \frac{\partial}{\partial x_j}+x_j \frac{\partial}{\partial x_i}$   & $J_{ij}$ &rotations\\
\hline 
(t $\to$ t + $\epsilon$) & $\frac{\partial}{\partial t}$ & $H$ & time  translation \\
\hline 
$\Big(t \to (1 - 2\epsilon)t, \,  x_i \to (1 - 2\ell \epsilon)x_i \Big)$ &$-2t\frac{\partial}{\partial t}-2 \ell x_i \frac{\partial}{\partial x_{i}}$ & $D$ &dilatation\\
\hline
$\Big(t \to (1 + \epsilon t)t, \,  x_i \to (1 + 2\ell \epsilon t)x_i \Big)$   &$t^2\frac{\partial}{\partial t} +2 \ell t x_i \frac{\partial}{\partial x_{i}}$ & $C$ &conformal transformation\\
\hline
$ (x_i \to x_i + \epsilon)$ & $ \frac{\partial}{\partial x_{i}}$ & $P_{0,i}$ &space  translation\\
$(x_i \to x_i - \epsilon t)$& $-t\frac{\partial}{\partial x_{i}}$ & $P_{1,i}$ &Galilean  boost \\
$(x_i \to x_i + \epsilon t^2)$ & $  t^2\frac{\partial}{\partial x_{i}}$ & $P_{2,i}$ &acceleration\\
$\Big(x_i \to x_i + \epsilon (-t)^n\Big)$ & $  (-t)^n\frac{\partial}{\partial x_{i}}$  & $P_{n,i}, n > 2 $&higher  derivative  transformation\\
\hline
    \end{tabular}\label{ta1b}
}
\end{table}
 The above generators satisfy the following non-trivial commutation relations 
 \begin{eqnarray}
[D,H]&=& 2H,   \,\,[D,C]= -2C, \,\,[C, H]=D, \,\,\label{CG1}
\qquad\qquad\qquad \\
\,\,[H,P_{n,i}]&=&-nP_ {n-1,i},\,\,[D,P_{n,i}]=2(\ell - n)P_ {n,i}, \,\,[C,P_{n,i}]=(2\ell - n)P_{n+1,i},\label{CG2}
\qquad\qquad\qquad \\
\,\,[J_{ij}, P_{n,k} ]&=&\delta_{ik}P_{n,j} - \delta_{jk}P_{n,i},  \,\,
[J_{ij}, J_{k\ell}] =\delta_{ik}J_{j\ell} + \delta_{j \ell}J_{ik} - \delta_{i \ell}J_{jk} - \delta_{jk}J_{i \ell}. \label{CG3}
\end{eqnarray}
 The conformal Galilei algebra with central extension, denoted
 $\mathfrak{\Hat{g}}_\ell(d)$,  has two types of central extension\cite{Mar10}:\\
 (1)  A central extension exists for any dimension of spacetime and  $\ell \in  \mathbb{N} +\frac{1}{2}$
\begin{eqnarray}
 [P_{m,i},P_{n,j}]= \delta _{ij}  \delta _{m+n,2\ell} I_{m}  M ,\quad\qquad I_{m}= (-1)^{m+\ell+1/2}  (2\ell -m)! \,\ m!, \label{CG4}
\end{eqnarray}
where $M$ is central. \\
(2) The so-called exotic extension, which exists only for d=2 and integer values of $\ell \in \mathbb{N}$
\begin{eqnarray}
[P_{m,i},P_{n,j}]= \epsilon _{ij} \delta _{m+n,2\ell} I_{m} \Theta,\quad\qquad \,\,\,\,\,\,\,\,\,\,\,\,\ I_{m}= (-1)^m (2\ell -m)! \,\ m!.\label{CG5}
\end{eqnarray}
Here $\Theta$ is central and $\epsilon _{ij}$ is the  antisymmetric tensor with
$\epsilon _{12}=1$. Notice that the dimension  of the conformal Galilei algebra
involving central extension  $\mathfrak{\Hat{g}}_\ell(d)$ is given by $2d\ell+\frac{d(d+1)}{2}+4$.

\section{Conformal Galilei algebra with $d=1$ and $\ell\geq \frac{3}{2}$}

First recall the formula by Beltrametti and Blasi \cite{BelBla66} for the
number of {\em generalised} Casimir operators of a Lie algebra $\mathfrak{g}$,
\begin{equation}
	\mbox{dim}(\mathfrak{g}) - \mbox{rank}(C(\mathfrak{g})),
	\label{bbformula}
\end{equation}
where $C(\mathfrak{g})$ denotes the commutator table of $\mathfrak{g}$ treated
as a numerical matrix with generic entries. In the cases below, we find
this number of polynomial Casimir operators.

For $d=1$, the Beltrametti-Blasi formula predicts that there are two
functionally independent Casimir operators for $\mathfrak{\Hat{g}}_\ell(1)$.
One is clearly the central element $M$. The other Casimir operator will be
constructed via the algorithm, the details of which
can be found in \cite{Als17}. We do not repeat the technical details here, but
will refer to the steps (A), (B), ... , (I) as presented in that article.

Following the algorithm, we need to first establish the artificial relative dimensions of
the basis elements, a type of grading of the Lie algebra in a certain basis, as
discussed in \cite{Als17}. We introduce artificial relative dimensions $\alpha$ and
$\beta$ and express the dimensions of all generators in a consistent way with
the commutation relations (\ref{CG1}),(\ref{CG2}) and (\ref{CG4}).
This leads to the following:
\begin{eqnarray}
[P_ {n}]&=&{\alpha^{\frac{2\ell-2n}{2\ell}}}{ \beta^{\frac{n}{2\ell}}},
[H]={\alpha^{\frac{2}{2\ell}}}/{ \beta^{\frac{1}{2\ell}}} ,\ \label{th7}  
\\
 \, [C]&=&{\beta^{\frac{1}{2\ell}}}/ {\alpha^{\frac{2}{2\ell}}}, \, [D]=1, \,  [M]=\beta, \nonumber 
\end{eqnarray} 
where $n = 0,1, 2, \ldots , 2\ell$. 

This step is useful in simplifying the process of taking linear combinations of
terms in steps (B) and (C) of the search algorithm.

We make an observation for low dimensional cases that the lowest order Casimir
operators has dimensions coinciding with $ M^2$, i.e., $[K]=\beta^2$. For
artificial relative dimensions $\alpha^x\beta^y$ for example, we simply denote
these in terms of the indices only, i.e. $(x,y)$. In such notation,
$[K]=(0,2)$.

For any half-odd integer $\ell$, we then make the assumption  that
$[K]=(0,2)$. This greatly reduces number of terms in the expression for the
Casimir operator. We find that this assumption works in our method for the
conformal Galilei algebra cases.

Next, the algorithm requires the construction of a differential operator
realisation, as outlined in step (D).  Those realisations already have
been given in the literature (see \cite{Ai13} for more detail).  The triangular
decomposition of $\mathfrak{\Hat{g}}_\ell \equiv \mathfrak{\Hat{g}}_\ell(1)$  is given by 
 \begin{eqnarray*}
	 \mathfrak{\Hat{g}}_\ell=\mathfrak{\Hat{g}}^+_{\ell} \oplus
	 \mathfrak{\Hat{g}}^0_{\ell} \oplus  \mathfrak{\Hat{g}}^-_{\ell}, 
\end{eqnarray*}
where, 
\begin{eqnarray*}
	\mathfrak{\Hat{g}}^+_{\ell}=\langle H, P_0, P_1, . . .,
	P_{\ell-\frac{1}{2}}\rangle,  \quad\qquad
	\mathfrak{\Hat{g}}^0_{\ell}=\langle D,  M\rangle,\quad\qquad
	\mathfrak{\Hat{g}}^-_{\ell}=\langle C, P_{\ell+\frac{1}{2}}, P_{\ell+\frac{3}{2}}, . . ., P_{2\ell}\rangle,
\end{eqnarray*}
denoting $\langle\cdots\rangle$ for the span, and where $\delta$ is a free
parameter related to the conformal weight of $D$.
The realisations of $\mathfrak{\Hat{g}}^0_{\ell}$ are given by 
\begin{eqnarray}
D=\delta-2t\frac{\partial}{\partial t}-\sum\limits_{j=0}^{\ell-\frac{1}{2}}2(\ell-j)x_{j}\frac{\partial}{\partial x_{j}},  \quad\qquad M=m,
\end{eqnarray}
while the realisations of $\mathfrak{\Hat{g}}^+_{\ell}$ are represented as 
\begin{eqnarray}
H=-\frac{\partial}{\partial t},  \quad\qquad  P_k=-\sum\limits_{j=0}^{k} \left(
     \begin{array}{c}
       k \\
       j
     \end{array}
   \right)t^{k-j}\frac{\partial}{\partial x_{j}}, 
\end{eqnarray} 
 where  $k=0, 1, . . . , \ell-\frac{1}{2}$. The realisations of
 $\mathfrak{\Hat{g}}^-_{\ell}$ as 
\begin{eqnarray}
C&=&t D +t^2\frac{\partial}{\partial t}+\frac{m}{2}\Bigg[ \Big( \ell +\frac{1}{2} \Big)! \Bigg]^2x^2_{\ell-\frac{1}{2}}-\sum\limits_{j=0}^{\ell-\frac{3}{2}}(2\ell-j)x_{j}\frac{\partial}{\partial x_{j+1}},\\
 P_k&=&m\sum\limits_{j=2\ell-k}^{\ell-\frac{1}{2}}\left(
     \begin{array}{c}
       k \\
       2\ell-j  
     \end{array}
   \right)I_{2\ell-j} \, t^{k-2\ell+j} x_{j}-\sum\limits_{j=0}^{\ell-\frac{1}{2}} \left(
     \begin{array}{c}
       k \\
       j
     \end{array}
   \right)t^{k-j}\frac{\partial}{\partial x_{j}}, 
\end{eqnarray}
for  $k=\ell+\frac{1}{2}, \ldots , 2\ell$.

Note that the $\mathfrak{\Hat{g}}_\ell(1)$ algebra is  spanned by  $2 \ell +5$
generators. For our purpose, it is convenient to fix the ordering of generators
as
\begin{eqnarray}
\{ M, P_{0},P_{1}, \ldots, P_{\ell- \frac{3}{2}}, H, P_{\ell- \frac{1}{2}}, D,
	P_{\ell+ \frac{1}{2}}, C, P_{\ell+\frac{3}{2}}, \ldots, P_{2\ell} \}. \label{ord=1}
\end{eqnarray}

We  proceed by applying the steps (E), (F) and (G) of the  search algorithm  to
construct the  Casimir operators. In fact, we have explicitly checked our
results for  $\mathfrak{\Hat{g}}_\ell(1)$ and  values
$\ell=\frac{3}{2},\frac{5}{2},. . . ,\frac{17}{2}$. Our algorithm has found
Casimir operators with the same artificial relative dimension as the central
element $M^2$, i.e. (0, 2).  In addition, this algorithm shows that there is
only one candidate
Casimir operator in the case of arbitrary $\mathfrak{\Hat{g}}_\ell(1)$ (in
reference to step (H) of the algorithm). 

Before we present the output of the algorithm, we need to first introduce a
useful lemma that makes use of an algebra anti-automorphism, which reduces a large amount of
calculations.

\subsection{Utility of an algebra anti-automorphism for  $d=1$} 
The aim of this section is to prove an assertion concerning an algebra anti-automorphism for  $d=1$. 
In the context of conformal Galilei algebras, the  following lemma shows that we may only need to consider verifying the commutator of the proposed Casimir operator with a subset of  basis vectors. Specifically for the case $d=1$, we define the involution anti-automorphism of the conformal Galilei algebra as\cite{Ai12}
\begin{eqnarray}
\omega(H)=C, \,\,\ \omega(C)=H, \,\,\ \omega(D)=D, \,\,\ \omega(M)=M, \,\,\ \omega(P_{n})=P_ {2\ell-n}.\label{omg1} 
\end{eqnarray}
\begin{lemma} \label{lem111}
\begin{eqnarray}
\text{If}  \,\ \omega(K)=K \, \text{and}\,\ [K, H]=0=[K,P_ {2\ell}], \,
	\text{then}\, [K, X]=0, \,\,\, \forall X \in \mathfrak{\Hat{g}}_{\ell}(1). \label{lem1}
\end{eqnarray}
\end{lemma}
 {\large \bf Proof:}
The proof is straightforward, using the commutation relations (\ref{CG1}),(\ref{CG2}),(\ref{CG4}) and the Jacobi identity.\\

The main results are summarised below, corresponding to different cases.

\subsubsection{$\mathfrak{\Hat{g}}_\frac{3}{2}(1)$ }\label{3/2} 

As already discussed, for this case the Beltrametti-Blasi formula
(\ref{bbformula}) implies that there exist two Casimir
operators for $\mathfrak{\Hat{g}}_\frac{3}{2}(1)$. One is clearly the central
element $M$. By applying the algorithm, the search for a quartic Casimir
operator with artificial relative dimensions (0, 2) was found
\begin{eqnarray}
K&=& -6 M^2 \,D +  M^2 \,D^{2} -4 M^2\, H\, C-\frac{7}{2} MP_{0}P_{3}
+ \frac{5}{2} MP_{1}P_{2}- 2M ( H P_{1}P_{3}+  P_{0}P_{2}C )  \nonumber\\
&+&\ 2 M( H P^{2}_{2}+ P^{2}_{1}C ) 
+   M  P_{0}DP_{3}
-M  P_{1}DP_{2}+ \frac{1}{4} P^{2}_{0}P^{2}_{3}-\frac{3}{2}P_{0}P_{1}
P_{2}P_{3}  - \frac{3}{4} P^{2}_{1}P^{2}_{2}\nonumber\\
&+&\  ( P^{3}_{1}P_{3}+  P_{0}P^{3}_{2}).\label{Ra1} 
\end{eqnarray} 
To check the expression $K$ in (\ref{Ra1}) is a Casimir operator, we employ the result of Lemma \ref{lem111}. It is clear that using the involution anti-automorphism in (\ref{omg1}), it straightforward to confirm that $\omega(K)=K$. It only remains to check the following relations  
\begin{eqnarray*}
 [K, H]=0=[K,P_ {3}].\label{Ra3} 
\end{eqnarray*}
Without including all the details of the calculation, we have that

\begin{eqnarray*}
[K,H]&=&   ( {-12+ 4  + 8}) {M^2H} +( {4-4 }) {M^2HD} + {(\frac{-21}{2}+\frac{5}{2}-2-2+12 )} {M P_{0}P_{2}} 
 \nonumber\\
&+&\   {( 5 - 2  -3 )}  {M P^2_{1}} +   {(2-2 )}  {MP_{0}H P_{3}}  +  {(-6+8-2 )}  {MH P_{1}P_{2}} + {(3- 3  )} {P^3_{1}P_{2}} 
\nonumber\\
&+&\   {(3-2-1 )}  {M P_{0}DP_{2}}  + {(2-2 )} {M P^2_{1}D }  +  {(4-4 )} {M P_{0}P_{1}C}+ {(\frac{6}{4} -\frac{3}{2}  )} {P^2_{0}P_{2}P_{3}}
 \nonumber\\
 &+&\  {(-\frac{9}{2}- \frac{3}{2}+6  )}  { P_{0}P_{1}P^2_{2}}  + {(3- 3  )}{P_{0}P^2_{1}P_{3}}  \nonumber\\
 &=&\  0.\label{Ra4} 
\end{eqnarray*}
Also, 
\begin{eqnarray*}
[K,P_ {3}]&=&( 18-9 -21+12) M^2 P_{3} + ( -6 +6 )  M^2 DP_{3}+  ( 12-12 )    M^2 P_{2}C  \nonumber \\
 &+&\  (6 +3 -9 )    M P_{1}P_{2} P_{3}+( -6 +6  )  M P^3_{2}+ (-3 +3 )  M P_{0}P^2_{3}\nonumber\\
 &=&\  0. 
\end{eqnarray*}
It is easily verified that the conditions of Lemma \ref{lem111} are satisfied,
and hence that  $K$ in equation (\ref{Ra1}) is indeed a Casimir operator of
$\mathfrak{\Hat{g}}_\frac{3}{2}(1)$.  

It is worth remarking that the Casimir operator $K$ in (\ref{Ra1}) has two
types of terms. One type is self-conjugate under $\omega$ (e.g.
$\omega(M P_0 P_3)=M P_0 P_3$), and the other type involves a conjugate pair under
$\omega$ (e.g. $\omega(M H P_1 P_3) = M P_0 P_2 C$).

 \subsubsection{$\mathfrak{\Hat{g}}_\frac{5}{2}(1)$}\label{5/2}

 Repeating the same process with the case $\ell=\frac{5}{2} $, there are still
 two functionally independent Casimir operators. One is clearly the central
 element $M$.  By applying the algorithm, a quartic Casimir operator was found with artificial relative dimensions (0, 2) as follows:
\begin{eqnarray}
K&=& 132 M^2 D -12  M^2 D^{2} +48  M^2 H C - MP_{0}P_{5}
+ 28 MP_{1}P_{4} -10 MP_{2}P_{3} -2 M ( P_{1}HP_{5}
 \nonumber \\
&+&\ P_{0}C P_{4}) + 8 M ( H P_{2}P_{4}+  P_{1}P_{3}C )-6 M (H P^{2}_{3} + P^{2}_{2}C) + M  P_{0}DP_{5} -3 M  P_{1}DP_{4}
\nonumber \\
&+&\  2 M  P_{2}DP_{3} - \frac{1}{48}P^{2}_{0}P^{2}_{5}+ \frac{5}{24}P_{0}P_{1}
P_{4}P_{5}-\frac{1}{12} P_{0}P_{2}P_{3}P_{5}-\frac{3}{16}P^{2}_{1}P^{2}_{4}+  \frac{19}{12}P_{1}P_{2}P_{3}P_{4} 
\nonumber \\
&+&\ \frac{2}{3} P^{2}_{2} P^{2}_{3} -\frac{1}{3} (P^{2}_{1} P_{3}P_{5}+ P_{0}P_{2} P^{2}_{4})+ \frac{1}{4}(P_{1} P^{2}_{2} P_{5}+ P_{0} P^{2}_{3}P_{4})- (   P^{3}_{2} P_{4}+  P_{1} P^{3}_{3}).  \label{Ra6}  
\end{eqnarray} 

Similar to the case of $\mathfrak{\Hat{g}}_\frac{3}{2}(1)$ in the previous
subsection \ref{3/2}, we employ the result of Lemma \ref{lem111}. By using the
involution anti-automorphism in (\ref{omg1}), it is obvious that $\omega(K)=K$.
Now, in order to see the remaining conditions of the Lemma \ref{lem111} are
satisfied,  we need to compute commutators of the Casimir operator $K$  in (\ref{Ra6}) with the
generators $H$  and  $P_ {5}$. A straightforward computation shows that
\begin{eqnarray*}
[K, H]=0=[K,P_ {5}].
\end{eqnarray*}

All the conditions for the Lemma \ref{lem111} are satisfied. Therefore, the
expression $K$ in (\ref{Ra6}) is indeed a Casimir operator of
$\mathfrak{\Hat{g}}_\frac{5}{2}(1)$. 

\subsubsection{$\mathfrak{\Hat{g}}_\ell(1)$, $\ell\geq \frac52$}

Using similar methods to subsections \ref{3/2} and \ref{5/2} we have explicitly
checked the Casimir operators for half-odd integer values of $\ell$ up to
$\frac{17}{2}$. It is clear from our computation that the case of
$\ell=\frac{3}{2}$ is a special case. Extending to arbitrary half-odd integer
value sof $\ell$, we summarise the results in the
following theorem, the proof of which follows similar calculations to those
already presented in subsections \ref{3/2} and \ref{5/2}. 
\begin{thm}
For $\ell \geq\frac{5}{2}$, $\mathfrak{\Hat{g}}_\ell(1)$ has a quartic Casimir operator of the form 
\begin{eqnarray}
K&=& \alpha  \, M^2 \,D +\beta \, M^2 \,D^{2} +\gamma \, M^2\, H\, C + \psi_{\ell- \frac{3}{2}}M \Bigg(  H P_{\ell - \frac{1}{2}}P_{\ell +\frac{3}{2}}+ P_{\ell - \frac{3}{2}} \ P_{\ell + \frac{1}{2}} C\Bigg) \nonumber \\
 &&\  + \psi_{\ell- \frac{1}{2}}M \Bigg(  H P^2_{\ell + \frac{1}{2}} + P^2_{\ell - \frac{1}{2}}   C\Bigg)+ \sum\limits_{i=0}^{\ell-\frac{1}{2}}\varphi_{i}M  A_{i}+ \sum\limits_{i=0}^{\ell-\frac{5}{2}} \psi_{i}M \Bigg(B_{i}+ C_{i}\Bigg)\nonumber \\
 &&\ + \sum\limits_{i=0}^{\ell-\frac{1}{2}}  \theta_{i}M D_{i}+ \sum\limits_{i=0}^{\ell-\frac{1}{2}}   \sum\limits_{j=i}^{\ell-\frac{1}{2}}  \tau_{ij} X_{ij} +   
\sum\limits_{i=0}^{\ell-\frac{3}{2}}   \sum\limits_{j=i}^{\ell-\frac{3}{2}} \lambda_{ij} Y_{ij},\label{TR1}   
\end{eqnarray} where, 
\begin{eqnarray*}
 &&A_{i}=P_{i}P_{2\ell-i}, \, B_{i}=  P_{i+1}HP_{2\ell-i}, \, C_{i} =P_{i} \ C \ P_{2\ell-i-1}, \,  D_{i} = P_{i}D P_{2\ell-i},  
 \\&& 
  X_{ij} = \  P_{i} \ P_{j} \ P_{2\ell-j} \ P_{2\ell-i}  , \,\,\,\  Y_{ij}= P_{i+1} \ P_{j+1} P_{2\ell-j-2} P_{2\ell-i}+ P_{i} P_{j+2} \ P_{2\ell-j-1} P_{2\ell-i-1},    
\end{eqnarray*}
and $\alpha,$ $\beta,$ $\gamma,$ $\psi_{\ell- \frac{3}{2}},$ $\psi_{\ell-
	\frac{1}{2}},$ $\varphi_{i},$ $\psi_{i},$ $\theta_{i},$ $\tau_{ij},$
	and $\lambda_{ij}$ are constants given by 
$$
 \alpha = \Big[\frac{1}{2}  (-1)^{\ell+\frac{3}{2}}(\ell+\frac{1}{2})^{2} (2\ell-1)!  -(-1)^{\ell+\frac{1}{2}}(2\ell-1)!  \Big] , \quad 
\beta =\frac{1}{2}  (-1)^{\ell+\frac{1}{2}} (2\ell-1)!   , 
$$
$$
\gamma =
-2 (-1)^{\ell+\frac{1}{2}} (2\ell-1)!  , 
 \quad 
  \psi_{\ell-\frac{1}{2}} =\frac{ -(-1)^{\ell+\frac{3}{2}} (2\ell-1)! }{\Big [(\ell-\frac{1}{2})!\Big]^2}  ,\qquad
$$
$$
  \psi_{i} =
  \frac{  2(-1)^{2\ell-i+2}\,\,(2\ell-1)! }{(i)!\,(2\ell-i-1)!} ,\,\,\,\,  i=
	0,\ldots, \ell-\frac{3}{2},\quad
 \varphi_{\ell-\frac{3}{2}}=
  \frac{  -7(-1)^{\ell+\frac{1}{2}}\,(\ell-\frac{1}{2})\,(2\ell-1)! }{2(\ell-\frac{3}{2})!\,(\ell+\frac{1}{2})!} , 
$$
$$
     \varphi_{\ell-\frac{1}{2}}=
    \frac{  (-1)^{\ell+\frac{1}{2}}\,\,(\ell+\frac{1}{2})\Big[ 5+4\ell(\ell+1)  \Big]  (2\ell-1)!\,\,   }{ 8\Big [(\ell+\frac{1}{2})!\Big]^2},
$$
$$
       \varphi_{i}= \frac{ (-1)^{i} \Big[ (-1)^{2\ell+1}\,\,i\,\,
	(2\ell+1)^2+\ell \Big(7-4\ell(\ell-1)\Big)  \Big]  (2\ell-1)!\,\,   }{
		4(i)!\,\, (2\ell-i)!},\,\,\,\,  i= 0,\ldots, \ell-\frac{5}{2}, 
$$
$$
  \theta_{i}=\frac{  (-1)^{2\ell-i+1}\,\,(2\ell-2i)\,\,(2\ell-1)!
	}{(i)!\,(2\ell-i)!} ,\,\,\,\,  i= 0,\ldots, \ell-\frac{1}{2}, 
$$
$$
  \tau_{\ell-\frac{1}{2}\ell-\frac{1}{2}} = \frac{  -(-1)^{\ell+\frac{1}{2}}\,(\ell+\frac{3}{2})\,(2\ell-1)! }{2(\ell+\frac{1}{2})\,(\ell-\frac{3}{2})!\,(\ell+\frac{1}{2})!\Big [(\ell+\frac{1}{2})!\Big]^2},\,\,\,\, 
$$
$$
  \tau_{ii+1}= \frac{  4(-1)^{\frac{1}{2}-\ell}\,\, \Big[
	  \frac{1}{2}(-1)^{2(\ell+i)}i \big(i-1-2\ell\big)+\big(i-\ell \big)^2
	  \Big] (2\ell-1)! }{\,(i+1) \, (2\ell-i)!\, (2\ell-i-1)!\,
	  \big(i!\big)^2},\,\,\,\,  i= 0,\ldots, \ell-\frac{3}{2},
$$
$$
  \tau_{ii}=  \frac{  -2(-1)^{\frac{1}{2}-\ell}\,\, \big(i-\ell \big)^2 \,
	(2\ell-1)! }{\big(i!\big)^2 \,  \big[(2\ell-i)!\big]^2},\,\,\,\,  i=
	0,\ldots, \ell-\frac{3}{2}, 
$$
$$
     \tau_{ij}=  \frac{  4(-1)^{\ell+i-j+\frac{5}{2}}\,\, \Big[ (-1)^{2j}i-\ell
     \Big] (j-\ell)(2\ell-1)! }{\,i! \,(j)!\, (2\ell-i)!\, (2\ell-j)!},\,\,\,\,
     i= 0,\ldots, \ell-\frac{5}{2},\,\,\,\, j=i+2,\ldots, \ell-\frac{1}{2},
$$
$$
      \lambda_{i\ell-\frac{3}{2} }= 
  \frac{  (-1)^{2\ell+i-1}\,\,(2\ell-1)!}{i!\,(2\ell-i-1)!\,\Big
  [(\ell-\frac{1}{2})!\Big]^2 },\,\,\,\,  i= 0,\ldots, \ell-\frac{3}{2},
$$
$$
  \lambda_{ij}= 
  \frac{  2(-1)^{\ell+i+j+\frac{1}{2}}\,\,(2\ell-1)!
  }{i!\,(j+1)!\,(2\ell-i-1)!\,(2\ell-j-2)!},\,\,\,\,  i= 0,\ldots,
  \ell-\frac{5}{2},\,\, j= i,\ldots, \ell-\frac{5}{2}.
$$
\end{thm}

\section{Conformal Galilei algebras with $d=2$ and $\ell\geq 1$}

Here we show that the algorithm also applies to the  construction of Casimir
operators for the conformal Galilei algebra with the exotic central extension
for $d=2$. Let us denote this algebra by $\mathfrak{\Hat{g}}_{\ell}(2)$, where
$\ell \in \mathbb Z_+$. Previously, these Casimir operators of
$\mathfrak{\Hat{g}}_{\ell}(2)$ have appeared in \cite{An12} and \cite{Luk7},
and our results are consistent with these works. 
 
A vector field realisation of the Conformal Galilei algebras with  central
extension has been found for $d=2$ and arbitrary values of  $\ell$\cite{Ai13}.
This algebra has triangular decomposition which can be written as a direct sum
of the vector space $\mathfrak{\Hat{g}}_\ell=\mathfrak{\Hat{g}}^+_{\ell} \oplus
\mathfrak{\Hat{g}}^0_{\ell} \oplus  \mathfrak{\Hat{g}}^-_{\ell} $, where
 % }
\begin{eqnarray*}
	&&\mathfrak{\Hat{g}}^+_{\ell}=\langle H,Q_{\ell} ,Q_{n},P_{n} \rangle,
	\qquad n=0,1,\ldots,\ell-1, \\&& \mathfrak{\Hat{g}}^0_{\ell}=\langle D, J,  \Theta \rangle , 
	\\&& \mathfrak{\Hat{g}}^-_{\ell}=\langle C, P_{\ell} ,Q_{n},P_{n}
	\rangle, \qquad   n=\ell+ 1,\ell+2, \ldots,2\ell.  
\end{eqnarray*} 

The following non-trivial commutation relations of the conformal Galilei
algebra $\mathfrak{\Hat{g}}_\ell(2)$ are given by
\begin{eqnarray}
&[D, H]&= 2H,  \quad   [D,C]=-2C ,  \quad [C,H]=D ,\label{g1}
\\&
[H,Q_{n}]&=-nQ_{n-1},          \quad [H,P_{n}]=-nP_{n-1} , \,\label{g2}
\\&
[D,Q_{n}]&=2(\ell-n)Q_{n},    \quad [D,P_{n}]=2(\ell-n)P_{n},\label{g3}                                                                                   
\\& 
[C,Q_{n}]&=(2\ell-n)Q_{n+1},   \quad [C,P_{n}]=(2\ell-n)P_{n+1},\label{g4}
\\&
[J,Q_{n}]&= Q_{n},\,\, [J,P_{n}]=-P_{n},\quad [Q_ {m},P_{n}]= \delta _{m+n,2\ell} I_m \Theta \label{g5}.                                                 
\end{eqnarray}
Note that  $I_m$ is given in   (\ref{CG5}).

Now, we work through the algorithm to produce the
lowest order Casimir operators. First of all, the algorithm
requires the use of differential operator realisations, described in step (D)
of \cite{Als17}. We use the realisations given in \cite{Ai13}.

Explicitly, the realisations of $\mathfrak{\Hat{g}}^+_{\ell}$  are given by 
\begin{eqnarray*}
H&=-\frac{\partial}{\partial t}, \quad  Q_{n} =-\sum\limits_{k=0}^{n} \left(
     \begin{array}{c}
       n \\
       k
     \end{array}
   \right)t^{k}\frac{\partial}{\partial x_{n-k}} , \quad
 P_{n} =-\sum\limits_{k=0}^{n} \left(
     \begin{array}{c}
       n \\
       k
     \end{array}
   \right)t^{k}\frac{\partial}{\partial y_{n-k}},
\end{eqnarray*}
for $n=0,1,\ldots,\ell-1$. For $\mathfrak{\Hat{g}}^0_{\ell}$, we have 
\begin{eqnarray*}
D&=&\delta-2t\frac{\partial}{\partial t}-\sum\limits_{n=0}^{\ell-1}2(\ell-n) \bigg(x_{n}\frac{\partial}{\partial x_{n}}+y_{n}\frac{\partial}{\partial y_{n}}\bigg), \\ 
 J&=&r-\sum\limits_{n=0}^{\ell} x_{n}\frac{\partial}{\partial x_{n}}+\sum\limits_{n=0}^{\ell-1} y_{n}\frac{\partial}{\partial y_{n}}, \quad \Theta=- \theta ,
\end{eqnarray*} \\
where $\delta$ is a free parameter, and the realisations of
$\mathfrak{\Hat{g}}^-_{\ell}$ that we use are given by 
\begin{eqnarray*}
C&=t D +t^2\frac{\partial}{\partial t}-\ell I_{\ell+1} \theta x_{\ell}y_{\ell-1}-\sum\limits_{n=0}^{\ell-1}(2\ell-n)x_{n}\frac{\partial}{\partial x_{n+1}}-\sum\limits_{n=0}^{\ell-2}(2\ell-n)y_{n}\frac{\partial}{\partial y_{n+1}},\\
 &\ Q_{n}=- \theta \sum\limits_{k=0}^{n-\ell-1}\left(
     \begin{array}{c}
       n \\
      k 
     \end{array}
   \right)I_{n-k} \, t^{k}{ y_{2\ell-n+k}}-\sum\limits_{k=n-\ell}^{n} \left(
     \begin{array}{c}
       n \\
      k
     \end{array}
   \right)t^{k}\frac{\partial}{\partial x_{n-k}}, \\ 
    &\ P_{n}=\theta \sum\limits_{k=0}^{n-\ell}\left(
     \begin{array}{c}
       n \\
      k 
     \end{array}
   \right)I_{n-k} \, t^{k} x_{2\ell-n+k}- \sum\limits_{k=n-\ell+1}^{n} \left(
     \begin{array}{c}
       n \\
      k
     \end{array}
   \right)t^{k}\frac{\partial}{\partial y_{n-k}}.   
\end{eqnarray*}

 where  $n=\ell+ 1,\ell+2, \ldots,2\ell$.
Note that $\mathfrak{\Hat{g}}_\ell(2)$ is spanned by  $4\ell +7$ elements. For
our purpose, it is convenient to fix the ordering of those generators within
elements of the enveloping algebra as
\begin{eqnarray}
\{ \Theta , Q_{0},P_{0},Q_{1},P_{1}, \ldots, Q_{\ell- 1},P_{\ell- 1}, H,D,J,
	Q_{\ell},P_{\ell},C, Q_{\ell+ 1},P_{\ell+ 1}, \ldots, Q_{2\ell},P_{2\ell} \}. \label{fd2}
\end{eqnarray}

From the Lie bracket given in (\ref{g1}-\ref{g5}), we are able to introduce
artificial relative dimensions $\alpha$ , $\beta$ and $\gamma$ and express the
dimensions of all generators in a consistent way with the commutation
relations. This leads to the following:
\begin{eqnarray}
[P_ {n}]&=& \frac{\alpha}{\gamma^{n}},\quad [Q_ {n}]= \frac{\beta}{\alpha} \gamma^{2 \ell - n},\ [C]=\frac{1}{\gamma},  \nonumber 
\\
 \,  [H]&=&\gamma, \, [D]=[J]=1, \,  [\Theta ]=\beta, \label{th155} 
\end{eqnarray} 
where $n = 0,1, 2, . . . , 2\ell$. As before, we adopt the notation using only
indices, so that if an element $A$ has artificial relative dimension
$[A]=\alpha^x\beta^y\gamma^z$, we simply write the triple $[A]=(x,y,z)$.

For  the conformal Galilei algebra $\mathfrak{\Hat{g}}_\ell(2)$, we make the
assumption (actually based on calculations for specific cases) that the lowest order Casimir operator has artificial relative
dimensions the same as the central element $\Theta$, i.e. (0,1,0) and $\Theta^2$
( artificial relative dimensions (0,2,0)). This greatly reduces number of terms
in the expression for the Casimir operator. We find that this assumption works
in our method for the conformal Galilei algebra cases.

We apply the steps (E), (F) and (G) of the search algorithm by using
Mathematica software in order to solve the corresponding linear algebraic equations. In the
case $d=2$, the algorithm leads us to the candidate Casimir operators which
arise from step (H).

\subsection{Utility of an algebra anti-automorphism for  $d=2$ }

Before summarising the output of the algorithm, let us present a functional
lemma, analogous to Lemma \ref{lem111}. It turns out for the case $d=2$, only three
elements are needed to verify the relations of the Casimir operator. 
Define the map
\begin{eqnarray}
\omega(H)&=&C, \,\,\ \omega(C)=H, \,\,\ \omega(D)=D,\,\,\ \omega(J)=J, \,\,\
 \nn \\
\omega(\Theta)&=&\Theta, \,\,\ \omega(P_{n})=Q_ {2\ell-n} ,\,\ \omega(Q_{n})=P_
	{2\ell-n}.\label{omg2} 
\end{eqnarray}
We have the following result, the proof of which is straightforward computation
involving commutators and the Jacobi identity. 
\begin{lemma}\label{lem222}
\begin{eqnarray}
\text{If}\,\,\ \omega(K)=K  \, \text{and}\,\ [K, H]=[K,J]=[K,P_ {2\ell}]=[K,Q_
	{2\ell}]=0, \,\text{then} \, [K, X]=0, \,\, \forall X \in
	\mathfrak{\Hat{g}}_{\ell}(2).
\nn 
\end{eqnarray} 
 \end{lemma}
 
It is easily seen that the Beltrametti-Blasi formula (\ref{bbformula}) implies that there exist
three Casimir operators for $\mathfrak{\Hat{g}}_\ell(2)$. One is
obviously the exotic central extension $\Theta$. To obtain the other two Casimir
operators, we proceed by applying the search algorithm to $\mathfrak{\Hat{g}}_\ell(2)$.

In what follows, the let $K^{(n),\ell}_{(r,s,t)}$ denote 
a polynomial Casimir operator for $\mathfrak{\Hat{g}}_{\ell}(2)$ of degree $n$ and relative
dimension $(r,s, t)$.

\subsubsection{The $\mathfrak{\Hat{g}}_{1}(2)$  case }\label{ECGA1}

Using the algorithm to search for a quadratic Casimir operator of artificial relative dimensions
(0, 1,0) yields the following:
\begin{eqnarray}
&K^{(2),1}_{(0,1,0)}&=-\Theta J- \frac{1}{2} Q_{0} P_{2}- \frac{1}{2} P_{0} Q_{2}+ Q_{1}P_{1}.\label{F2}
\end{eqnarray}
Next, we seek a quartic Casimir operator of artificial relative dimensions
(0,2,0). The algorithm indicates two candidates which are not actual Casimir
operators, given by
 \begin{eqnarray*}
  K_a&=& -3 \Theta ^2D-6 \Theta ^2 J+\frac{1}{2} \Theta ^2 D^{2}-2 \Theta ^2 HC+2 \Theta ^2 J^2 - 4 \Theta  Q_{0} P_{2}+2 \Theta HP_{1}Q_{2} 
 \nonumber\\
 &-& 2 \Theta  HQ_{1}P_{2} -\Theta  P_{0} D Q_{2}+2 \Theta   P_{0}Q_{1}C +\Theta  Q_{0} D P_{2}-2 \Theta  Q_{0}P_{1}C +2 P_{0}  Q^2_{1} P_{2}
 \nonumber\\
 &+& 2 Q_{0}  P^2_{1} Q_{2}-2 Q^2_{1} P^2_{1}  -2 Q_{0}  P_{0}Q_{2} P_{2},
  \nonumber\\
   K_b&=&-6 \Theta ^2 J+2 \Theta ^2 J^2- 2 \Theta  P_{0} Q_{2}- 2 Q^2_{1} P^2_{1}- 2 Q_{0}P_{0}Q_{2} P_{2}- \frac{1}{2}Q^2_{0} P^2_{2} -\frac{1}{2}  P^2_{0} Q^2_{2} \nonumber\\
 &+&  2Q_{0} Q_{1}P_{1}P_{2}  + 2P_{0} Q_{1 }P_{1} Q_{2}. 
\end{eqnarray*}

Using the involution anti-automorphism given in (\ref{omg2}), we see that $\omega(K_a)=K_a$ and 
$\omega(K_b)=K_b$. 
Following step (I), we take a linear combination of the above candidates
 \begin{eqnarray*}
 K^{(4),1}_{(0,2,0)}=   \alpha  K_a + \beta  K_b,
 \end{eqnarray*}
  and impose the relations $ [K^{(4),1}_{(0,2,0)},
  H]=[K^{(4),1}_{(0,2,0)},J]=[K^{(4),1}_{(0,2,0)},P_ {2}]=[K^{(4),1}_{(0,2,0)},Q_
  {2}]=0.$ We find that
\begin{eqnarray*}
&K^{(4),1}_{(0,2,0)}&=  K_a-K_b .
\end{eqnarray*}
That is, 
\begin{eqnarray}
K^{(4),1}_{(0,2,0)}&=&  -3 \Theta^2 D +\frac{1}{2} \Theta^2 D^{2} -2 \Theta^2 H C+2 \Theta  \big (H P_{1}Q_{2}+ P_{0}Q_{1 }C\big) -2 \Theta  \big(H Q_{1}P_{2}+ Q_{0}P_{1 }C\big) \nonumber\\
&&\ -2\big( Q_{0} Q_{1} P_{1} P_{2}+ P_{0} Q_{1} P_{1} Q_{2} \big )+2 \Theta   P_{0} Q_{2}-4 \Theta   Q_{0} P_{2}-  \Theta  P_{0}  D  Q_{2}+ \Theta   Q_{0}  D  P_{2}\nonumber\\
&&\ +2 \big ( P_{0}Q^{2}_{1}P_{2} +Q_{0}P^{2}_{1}Q_{2} \big)- Q_{0}P_{0}Q_{2}P_{2}+\frac{1}{2} \big( P^{2}_{0} Q^{2}_{2}+ Q^{2}_{0} P^{2}_{2} \big).\label{exg1}
\end{eqnarray}
Hence, by Lemma \ref{lem222}, we have the Casimir operator for
$\mathfrak{\Hat{g}}_{1}(2)$.

\subsubsection{The $\mathfrak{\Hat{g}}_{2}(2)$  case }\label{ECGA2}

In this case we expect three Casimir operators, one of which is $\Theta$. Using
the algorithm, we find quadratic and
quartic Casimir operators. The quadratic Casimir operator of artificial
relative dimensions (0, 1,0) can be written as
\begin{eqnarray}
&K^{(2),2}_{(0,1,0)}&=-\Theta J- \frac{1}{24} Q_{0} P_{4}- \frac{1}{24} P_{0} Q_{4}+ \frac{1}{6}Q_{1}P_{3}+\frac{1}{6}P_{1}Q_{3}-\frac{1}{4}Q_{2}P_{2}.\label{F2b}
\end{eqnarray}
For the quartic Casimir operator of artificial relative dimensions (0, 2,0), the algorithm indicates two candidates as follows:
\begin{align}
K_a=& -54 \Theta ^2 D +3 \Theta ^2 D^{2}+12 \Theta ^2 D J -12 \Theta ^2 H C +5 \Theta   P_{0} Q_{4}-16  \Theta   P_{1} Q_{3}- \Theta   Q_{0} P_{4} 
\nonumber\\
&+20 \Theta  Q_{1} P_{3}  + 3 \Theta D Q_{2} P_{2} -6 \Theta  \big (H P_{2}Q_{3}+ P_{1}Q_{2}C\big)+6 \Theta  \big(H Q_{2}P_{3}+ Q_{1}P_{2}C\big) 
\nonumber\\
& +2 \Theta  \big (P_{1}HQ_{4}+ P_{0}CQ_{3} \big) -2 \Theta  \big( Q_{1}HP_{4}+ Q_{0}CP_{3}\big)  -\frac{1}{2}  \Theta P_{0} D Q_{4}+ \frac{3}{2} \Theta  Q_{0}  D  P_{4}
\nonumber\\
& -4 \Theta   Q_{1}  D  P_{3}+ 3 \big ( P_{1}Q^{2}_{2}P_{3} +Q_{1}P^{2}_{2}Q_{3} \big)+ \big(P_{0} P_{2}  Q^{2}_{3}+P^{2}_{1}  Q_{2} Q_{4}+ Q_{0} Q_{2} P^{2}_{3}  +Q^{2}_{1} P_{2} P_{4}\big)
\nonumber\\
&    +  \frac{1}{3} \big(  P_{0} Q_{1}  Q_{3} P_{4} +Q_{0}  P_{1} P_{3} Q_{4} \big) - \big ( P_{0}Q_{2}Q_{3}P_{3} +Q_{1}P_{1}P_{2}Q_{4} \big)+\frac{1}{12} \big( P^{2}_{0} Q^{2}_{4}+ Q^{2}_{0} P^{2}_{4} \big)
\nonumber\\
& - \big( Q_{0} P_{2}  Q_{3} P_{3} +Q_{1}  P_{1} Q_{2} P_{4}\big) + \frac{1}{3} \big( P^{2}_{1} Q^{2}_{3}+ Q^{2}_{1} P^{2}_{3} \big) -\frac{2}{3} Q_{1}P_{1}Q_{3}P_{3}-  \frac{1}{6} Q_{0}P_{0}Q_{4}P_{4}
\nonumber\\
&  -\frac{2}{3} \big( P_{0} P_{1} Q_{3}  Q_{4}+  Q_{0} Q_{1} P_{3} P_{4} \big)
+   \frac{1}{3} \big(  P_{0} Q_{1}  P_{3} Q_{4} +Q_{0}  P_{1} Q_{3} P_{4} \big) 
\nonumber\\
&
-3 \big(  P_{1} Q_{2}  P_{2} Q_{3}+Q_{1}  Q_{2}  P_{2}P_{3} \big), 
\end{align}
\begin{align}
 K_b=&-12 \Theta ^2 D + 12 \Theta ^2 D J +2 \Theta   P_{0} Q_{4}-4 \Theta   P_{1} Q_{3}+ 2 \Theta   Q_{0} P_{4}-4 \Theta  Q_{1} P_{3} +3  \Theta D Q_{2} P_{2}
 \nonumber\\
& + \frac{1}{2}  \Theta  P_{0} D Q_{4}-2  \Theta  P_{1} D Q_{3}  + \frac{1}{2} \Theta   Q_{0}  D  P_{4}   - 2 \Theta  Q_{1}  D  P_{3} \nonumber 
\end{align}
It turns out neither of these are Casimir operators, but behave as such when
restricted to the realisation. As in the $\mathfrak{\Hat{g}}_{1}(2)$  case in the previous section, by imposing the conditions of Lemma in \ref{lem222}, we find
\begin{eqnarray*}
&K^{(4),2}_{(0,2,0)}&= K_a-K_b.
\end{eqnarray*}
Specifically,
\begin{eqnarray}
K^{(4),2}_{(0,2,0)}&=& -42 \Theta ^2 D +3 \Theta ^2 D^{2} -12 \Theta ^2 H C+2 \Theta  \big (P_{1}HQ_{4}+ P_{0}CQ_{3} \big) -2 \Theta  \big( Q_{1}HP_{4}+ Q_{0}CP_{3}\big) \nonumber\\
\ && - 6 \Theta  \big (H P_{2}Q_{3}+ P_{1}Q_{2}C\big)+6 \Theta  \big(H Q_{2}P_{3}+ Q_{1}P_{2}C\big)-3 \big( Q_{1} Q_{2} P_{2} P_{3}+ P_{1} Q_{2} P_{2} Q_{3} \big )\nonumber\\
&&\ +3  \Theta   P_{0} Q_{4}-3 \Theta   Q_{0} P_{4}-12 \Theta  P_{1} Q_{3}+24  \Theta  Q_{1} P_{3}- \Theta  P_{0} D Q_{4}+ \Theta   Q_{0}  D  P_{4} \nonumber\\
&&\ +2 \Theta   P_{1}  D  Q_{3}- 2 \Theta   Q_{1}  D  P_{3}+\frac{1}{3} \big ( P_{0}Q_{1}Q_{3}P_{4} +Q_{0}P_{1}P_{3}Q_{4} \big)+ 3 \big ( P_{1}Q^{2}_{2}P_{3} +Q_{1}P^{2}_{2}Q_{3} \big)\nonumber\\
&&\  - \frac{1}{6} Q_{0}P_{0}Q_{4}P_{4}-\frac{2}{3} Q_{1}P_{1}Q_{3}P_{3}- \big ( P_{0}Q_{2}Q_{3}P_{3} +Q_{1}P_{1}P_{2}Q_{4} \big)+\frac{1}{12} \big( P^{2}_{0} Q^{2}_{4}+ Q^{2}_{0} P^{2}_{4} \big) \nonumber\\
&&\ -\frac{2}{3} \big( P_{0} P_{1} Q_{3}  Q_{4}+  Q_{0} Q_{1} P_{3} P_{4} \big)+\frac{1}{3} \big( P^{2}_{1} Q^{2}_{3}+ Q^{2}_{1} P^{2}_{3} \big)- \big( Q_{0} P_{2}  Q_{3} P_{3} +Q_{1}  P_{1} Q_{2} P_{4}\big)\nonumber\\
&&\  +  \frac{1}{3} \big(  P_{0} Q_{1}  P_{3} Q_{4} +Q_{0}  P_{1} Q_{3} P_{4} \big)+  \big(P_{0} P_{2}  Q^{2}_{3}+P^{2}_{1}  Q_{2} Q_{4}
+ Q_{0} Q_{2} P^{2}_{3}  +Q^{2}_{1} P_{2} P_{4}\big).\nonumber\\ \label{exg2}
\end{eqnarray}
It is easily verified that the conditions of Lemma  in \ref{lem222} are
satisfied, and hence that $K^{(4),2}_{(0,2,0)}$ is a  Casimir operator for
$\mathfrak{\Hat{g}}_{2}(2)$.

\subsubsection{The $\mathfrak{\Hat{g}}_{3}(2)$  case }\label{ECGA3}

At the risk of being monotonous, we look at one final value $\ell=3$ before
summarising in the general case. For the case $\ell=3$, we expect three functionally independent Casimir
operators, as in the previous cases $\ell=1$ and $\ell=2$. In applying our algorithm, a quadratic Casimir
operator  was found with artificial relative dimensions $(0, 1,0)$ as follows:
\begin{eqnarray}
K^{(2),3}_{(0,1,0)}&=&- \Theta J- \frac{1}{720} Q_{0} P_{6}- \frac{1}{720} P_{0} Q_{6}+ \frac{1}{120}Q_{1}P_{5}+\frac{1}{120}P_{1}Q_{5}-\frac{1}{48}Q_{2}P_{4}-\frac{1}{48}P_{2}Q_{4}\nonumber\\
 &+& \frac{1}{36}Q_{3}P_{3}.   \label{F2c}
\end{eqnarray}
A functionally independent quartic Casimir operator was also found with
artificial relative dimensions $(0, 2,0)$:
\begin{align}
K^{(4),3}_{(0,2,0)}&= -1560\Theta ^2 D +60 \Theta ^2 D^{2} -240 \Theta ^2 H C+8 \Theta   P_{0} Q_{6}-24 \Theta   P_{1} Q_{5}+60  \Theta  P_{2} Q_{4} -8 \Theta  Q_{0} P_{6}
\nonumber\\
 & +   24 \Theta   Q_{1} P_{5}-120 \Theta   Q_{2} P_{4}+ 2 \Theta   \big( P_{1} H Q_{6}+ P_{0} C Q_{5}\big)-10  \Theta   \big( P_{2} H Q_{5}+ P_{1} C Q_{4}\big) \nonumber\\
 &  +  20  \Theta \, \big( H  P_{3} Q_{4}+P_{2} Q_{3} C \big)-2  \Theta \, \big( Q_{1} H P_{6}+ Q_{0} C P_{5}\big)+ 10 \Theta \, \big( Q_{2} H P_{5}+ Q_{1} C P_{4}\big) \nonumber\\
 &- 20 \Theta \, \big(H  Q_{3} P_{4}  + Q_{2} P_{3} C \big)  - \Theta \,  P_{0}  D  Q_{6}+4 \Theta  \, P_{1}  D  Q_{5} -5 \Theta  \, P_{2}  D  Q_{4}  + \Theta \,  Q_{0}  D  P_{6}
  \nonumber\\
&- 4 \Theta \,  Q_{1}  D  P_{5} +5 \Theta \,  Q_{2}  D  P_{4}-\frac{5}{3} \big(  Q_{2}Q_{3} P_{3} P_{4} + P_{2}Q_{3} P_{3} Q_{4} \big) + \frac{1}{60} \big(  P_{0}Q_{1} Q_{5} P_{6}
 \nonumber\\
 &+  Q_{0}P_{1} P_{5} Q_{6} \big)+ \frac{5}{12} \big(  P_{1}Q_{2} Q_{4} P_{5}+ Q_{1}P_{2} P_{4} Q_{5} \big) +\frac{5}{3} \big(  P_{2}Q_{3} Q_{3} P_{4}+ Q_{2}P_{3} P_{3} Q_{4} \big) 
  \nonumber\\
 &- \frac{1}{120}  Q_{0} P_{0}  Q_{6}P_{6}-  \frac{2}{15}  Q_{1} P_{1}  Q_{5}P_{5}   - \frac{5}{24}  Q_{2} P_{2}  Q_{4}P_{4}-\frac{1}{12} \big(  P_{0}Q_{2} Q_{5} P_{5}+ Q_{1}P_{1} P_{4} Q_{5} \big)
 \nonumber\\
 &- \frac{5}{6} \big(  P_{1}Q_{3} Q_{4} P_{4}+ Q_{2}P_{2} P_{3} Q_{5} \big)  +\frac{1}{240} \big(   P^{2}_{0}  Q^{2}_{6}+ Q^{2}_{0} P^{2}_{6} \big)-\frac{1}{20} \big(  P_{0}P_{1}Q_{5} Q_{6} + Q_{0}Q_{1}P_{5} P_{6}  \big)
 \nonumber\\
&+ \frac{1}{24} \big(  P_{0}P_{2}Q_{4} Q_{6} + Q_{0}Q_{2}P_{4} P_{6}\big)+  \frac{1}{15} \big(   P^{2}_{1}  Q^{2}_{5}+ Q^{2}_{1} P^{2}_{5} \big)-\frac{7}{12} \big(  P_{1}P_{2}Q_{4} Q_{5} + Q_{1}Q_{2}P_{4} P_{5}\big)
\nonumber\\
 &+\frac{5}{48} \big(   P^{2}_{2}  Q^{2}_{4}+ Q^{2}_{2} P^{2}_{4} \big)     - \frac{1}{12} \big(  Q_{0}P_{2} Q_{5} P_{5}+ Q_{1}P_{1} Q_{4} P_{6}\big)+\frac{1}{6} \big(  Q_{0}P_{3} Q_{4} P_{5}+ Q_{1}P_{2} Q_{3} P_{6}\big)
 \nonumber\\
&-\frac{5}{6} \big(  Q_{1}P_{3} Q_{4} P_{4}+ Q_{2}P_{2} Q_{3} P_{5}\big)    +\frac{1}{30} \big( P_{0} Q_{1}  P_{5} Q_{6}+ Q_{0}P_{1} Q_{5} P_{6}\big)-\frac{1}{24} \big( P_{0} Q_{2}  P_{4} Q_{6}
 \nonumber\\
& + Q_{0}P_{2} Q_{4} P_{6}\big)+\frac{1}{6} \big( P_{1} Q_{2}  P_{4} Q_{5}+ Q_{1}P_{2} Q_{4} P_{5}\big)    +  \frac{1}{12} \big( P_{0}  P_{2} Q^{2}_{5}+ P^{2}_{1} Q_{4} Q_{6}+ Q_{0} Q_{2}P^{2}_{5}
 \nonumber\\
 & + Q^{2}_{1}P_{4} P_{6}\big)-\frac{1}{6} \big(  P_{0}P_{3} Q_{4}Q_{5} + P_{1}P_{2} Q_{3}Q_{6}+Q_{0}Q_{3} P_{4}P_{5}   +  Q_{1}Q_{2} P_{3}P_{6}\big)+ \frac{5}{6} \big(  P_{1}P_{3} Q^{2}_{4} 
 \nonumber\\
 &+ P^{2}_{2} Q_{3}Q_{5}+Q_{1}Q_{3} P^{2}_{4}+ Q^{2}_{2} P_{3}P_{5}\big)+\frac{1}{6} \big(  P_{0}Q_{3} P_{4}Q_{5} + P_{1} Q_{2} P_{3}Q_{6}\big).
 \end{align}
The form of $K^{(4),3}_{(0,2,0)}$ given above is slightly different to the
quartic Casimir operators given for the cases $\ell=1$ and $\ell=2$. It turns
out, however, that this form prevails for other values $\ell\geq 3$ as we see
in the next section.

\subsubsection{The $\mathfrak{\Hat{g}}_{\ell}(2)$  case }

Here we summarise the explicit expressions of the Casimir operators determined
by the algorithm for the
exotic conformal Galilei algebra in $d=2$ and arbitrary integer $\ell$. In
fact, this algebra has three functionally independent Casimir operators.
Clearly, the central extension $\Theta$ is  a Casimir operator. The others are
quadratic and quartic Casimir operators. The following expression of quadratic
Casimir operator of artificial relative dimension $(0, 1,0)$ was deduced from the
forms of the Casimir operators given in (\ref{F2}), (\ref{F2b}) and (\ref{F2c}),
but the proof is a straightforward but tedious exercise in commutation
relations.

\begin{thm}
For $d=2$ and $\ell \geq1$, the quadratic Casimir operator of the exotic conformal Galilei algebra has the form
\begin{eqnarray}
K^{(2),\ell}_{(0,1,0)}= -\Theta J+ \sum\limits_{m=0}^{\ell- 1}\frac{(-1)^{m+1}}{m! (2\ell- m)!} \Bigg(Q_{m} P_{2\ell- m}+ P_{m}Q_{2\ell- m}\Bigg)+(-1)^{\ell+ 1} \frac{1}{(\ell !)^{2}}Q_{\ell}P_{\ell}. \label{r2a}
\end{eqnarray} 
\end{thm}

For the quartic Casimir operator of the exotic conformal Galilei algebra
$\mathfrak{\Hat{g}}_{\ell}(2)$, where $\ell \geq 3$. It is clearly from our results
thta the form of the quartic Casimir operators for $\ell=1$ and $\ell=2$ are
special cases of $\mathfrak{\Hat{g}}_{\ell}(2)$. The following expression was
deduced from the forms of the Casimir operators corresponding to specific
values of $\ell$, which are shown to be
genuine Casimir operators by the tedious task of verifying commutation relations. 

\begin{thm}
For $d=2$  and arbitrary integer $\ell \geq3$, there is a quartic Casimir operator of artificial relative
dimension $(0, 2,0)$ given by
\begin{eqnarray}
&&K^{(4),\ell}_{(0,2,0)}= \alpha \, \Theta^2 \,D +\beta \, \Theta^2 \,D^{2} +\gamma \, \Theta^2\, H\, C + \psi_{\ell -1}\Theta \Big (H P_{\ell}Q_{\ell +1}+ P_{\ell -1}Q_{\ell }C\Big) \nonumber\\
&&\ + \psi^{'}_{\ell -1} \Theta \Big(H Q_{\ell}P_{\ell +1}+ Q_{\ell -1}P_{\ell }C\Big) +\delta_{\ell -1}\Big( Q_{\ell -1} \ Q_{\ell } \ P_{\ell} \ P_{\ell+1}+ P_{\ell -1} \ Q_{\ell } \ P_{\ell} \ Q_{\ell+1} \Big ) \nonumber\\
  &&\ + \sum\limits_{i=0}^{\ell- 1}\varphi_{i} \Theta A_{i} + \sum\limits_{i=0}^{\ell- 1}\varphi^{'}_{i}\Theta  A^{'}_{i} 
 +\sum\limits_{i=0}^{\ell- 2} \psi_{i}\Theta \Bigg(B_{i}+ C_{i}\Bigg) + \sum\limits_{i=0}^{\ell- 2} \psi^{'}_{i}\Theta \Bigg(B^{'}_{i}+ C^{'} _{i}\Bigg)
 \nonumber\\
  &&\ + \sum\limits_{i=0}^{\ell- 1}  \phi_{i}\Theta D_{i}+ \sum\limits_{i=0}^{\ell- 1}  \phi^{'}_{i}\Theta D^{'}_{i} + \sum\limits_{i=0}^{\ell- 1} \rho_{i} S_{i} + 
 \sum\limits_{i=0}^{\ell- 1} \eta_{i} R_{i} +  \sum\limits_{i=0}^{\ell- 2} \mu_{i} T_{i}+ \sum\limits_{i=0}^{\ell-1}   \sum\limits_{j=i}^{\ell-1}  \tau_{ij} X_{ij} \nonumber\\
&&\ +   \sum\limits_{i=0}^{\ell -2}   \sum\limits_{j=i}^{\ell - 2} \lambda_{ij} Y_{ij} +    \sum\limits_{i=0}^{\ell -2}   \sum\limits_{j=i}^{\ell - 2} \zeta_{ij} Z_{ij} +    \sum\limits_{i=0}^{\ell -2}   \sum\limits_{j=i}^{\ell - 2} \omega_{ij} U_{ij}+    \sum\limits_{i=0}^{\ell -3}   \sum\limits_{j=i}^{\ell - 3} \varepsilon_{ij} V_{ij},  \label{TRb} 
\end{eqnarray}
where    
$$
A_{i}=P_{i}Q_{2\ell-i}, \, A^{'}_{i}=Q_{i}P_{2\ell-i}, B_{i}=  P_{i+1}HQ_{2\ell-i}, \, B^{'}_{i}= Q_{i+1}HP_{2\ell-i}, 
$$
$$
  C_{i} =P_{i} C Q_{2\ell-i-1}, \, C^{'}_{i} = Q_{i}  C  P_{2\ell-i-1},
  D_{i} = P_{i}D Q_{2\ell-i}, D^{'}_{i} = Q_{i}D P_{2\ell-i}, 
$$
$$
 S_{i} = P_{i}Q_{i+1}Q_{2\ell-i-1}P_{2\ell-i} +Q_{i}P_{i+1}P_{2\ell-i-1}Q_{2\ell-i},\, R_{i} = Q_{i}P_{i}Q_{2\ell-i}P_{2\ell-i},
$$
$$
T_{i} =P_{i}Q_{i+2}Q_{2\ell-i-1}P_{2\ell-i-1} +Q_{i+1}P_{i+1}P_{2\ell-i-2}Q_{2\ell-i},
$$
$$
X_{ij} = \  P_{i} \ P_{j} \ Q_{2\ell-j} \ Q_{2\ell-i}+\  Q_{i} \ Q_{j} \ P_{2\ell-j} \ P_{2\ell-i}  ,
$$
$$
   Y_{ij}=Q_{i} P_{j+2}  Q_{2\ell-j-1} P_{2\ell-i-1} +Q_{i+1}  P_{j+1} Q_{2\ell-j-2} P_{2\ell-i},
$$
$$
  Z_{ij}=P_{i} Q_{j+1}  P_{2\ell-j-1} Q_{2\ell-i} +Q_{i}  P_{j+1} Q_{2\ell-j-1} P_{2\ell-i}, 
$$
$$
U_{ij}=P_{i} P_{j+2}  Q_{2\ell-j-1} Q_{2\ell-i-1} 
P_{i+1} P_{j+1} Q_{2\ell-j-2} Q_{2\ell-i}+ Q_{i} Q_{j+2} P_{2\ell-j-1} P_{2\ell-i-1} 
+ Q_{i+1} Q_{j+1} P_{2\ell-j-2} P_{2\ell-i}, 
$$
$$
V_{ij}=P_{i} Q_{j+3} \ P_{2\ell-j-2} Q_{2\ell-i-1} +P_{i+1} \ Q_{j+2} P_{2\ell-j-3} Q_{2\ell-i}, 
$$
and
$\alpha,$ $\beta,$ $\gamma,$ $\psi_{\ell -1},$ $\psi^{'}_{\ell -1},$ $
	\delta_{\ell -1},$ $\varphi_{i},$ $\varphi^{'}_{i},$ $\psi_{i},$
	$\psi^{'}_{i},$ $\phi_{i},$ $\phi^{'}_{i},$ $\rho_{i},$ $\eta_{i},$
	$\mu_{i},$ $\tau_{ij},$ $\lambda_{ij},$ $\zeta_{ij},$ $\omega_{ij},$
	$\varepsilon_{ij}$ are constants. The form of these are given by 
$$
 \alpha = \Big[  (-1)^{2\ell+1} \ell(\ell+1) (2\ell-1)!-(2\ell-1)!  \Big], \quad  \beta = \frac{1}{2}   (2\ell-1)!,
	$$
	$$
 \gamma = -2  (2\ell-1)! , \qquad \delta_{\ell-1} =\frac{ -2 (2\ell-1)! }{\Big [(\ell)!(\ell-1)!\Big]^2} , 
$$
$$
   \varphi_{i}= \frac{ -(-1)^{i} \Big[ 1+(-1)^{2\ell}(i-\ell)  \Big](\ell+1)
	(2\ell)! }{ (i)!\,\, (2\ell-i)!},\,\,\,\,  i= 0,\ldots, \ell-2,  
$$
$$
   \varphi^{'}_{i}= \frac{ (-1)^{i} \Big[ -1+(-1)^{2\ell}(\ell-i)
   \Big](\ell+1)  (2\ell)!}{ (i)!\,\, (2\ell-i)!},\,\,\,\,  i= 0,\ldots, \ell-2,
   $$
   $$
   \varphi_{\ell-1} =\frac{ -2(-1)^{\ell} (2\ell-1)! }{\Big [(\ell-1)!\Big]^2} ,\qquad   \varphi^{'}_{\ell-1} =\frac{ 4(-1)^{\ell} (2\ell-1)! }{\Big [(\ell-1)!\Big]^2},
    $$
    $$
    \psi_{i} = \frac{  2(-1)^{2\ell-i}\,\,(2\ell-1)!
    }{(i)!\,(2\ell-i-1)!},\quad   i= 0,\ldots, \ell-1, 
    $$
    $$
 \psi^{'}_{i} = \frac{ - 2(-1)^{2\ell-i}\,\ (2\ell-1)!
 }{(i)!(2\ell-i-1)!},\quad   i= 0,\ldots, \ell-1, 
 $$
 $$
  \phi_{i}=\frac{  -(-1)^{2\ell-i}\,\,(2\ell-2i)\,\,(2\ell-1)!
  }{(i)!\,(2\ell-i)!},\quad   i= 0,\ldots, \ell-1 ,
  $$
  $$
   \phi^{'}_{i}=\frac{  (-1)^{2\ell-i}\,\,(2\ell-2i)\,\,(2\ell-1)!
   }{(i)!\,(2\ell-i)!}, \quad i= 0,\ldots, \ell-1 
   $$
   $$
\rho_{i}=\frac{ -2 (-1)^{2\ell-2i-1}(2\ell-1)! }{ \Big [
	(i)!(2\ell-i-1)!\Big]^2},\quad  i= 0,\ldots, \ell-1 
$$
$$
   \mu_{i} =\frac{ 2 (-1)^{2\ell-2i-1}(2\ell-1)!
   }{(i)!(i+1)!(2\ell-i-1)!(2\ell-i-2)!}, \quad  i= 0,\ldots, \ell-2,
$$
$$
  \eta_{i}= \frac{  -4(-1)^{2\ell-2i}\,\, \Big[ (-1)^{2i}
  \big(1+i-\ell\big)-1\Big] \big(i-\ell \big)(2\ell-1)! }{ \Big [
	  (i)!(2\ell-i-1)!\Big]^2},\quad  i= 0,\ldots, \ell-1 
$$
$$
  \tau_{\ell-1\ell-1}= \frac{  -(2\ell-1)!\Big[ (\ell+2)\,(\ell-1)! (\ell)!- \ell  (\ell-2)! (\ell+1)! \Big] }{(\ell-2)!\,(\ell)!\,\Big [(\ell-1)!(\ell+1)!\Big]^2}, \quad \tau_{00}=\frac{ 1 }{2(2\ell-1)!},   
$$
$$
  \tau_{ii+1}= \frac{  2(-1)^{2\ell-2i}\,\, \Big[ i+i^2-2i\ell+2\ell^2 \Big]
  (i+1)(i-2\ell)(2\ell-1)! }{ \Big [ (i+1)!(2\ell-i)!\Big]^2},  \quad i=
  0,\ldots, \ell-3, 
$$
$$
    \tau_{i+1i+1}= \frac{  2(-1)^{2\ell-2i}\,\, \Big(i-\ell+1 \Big)^2
    (2\ell-1)! }{ \Big [ (i+1)!(2\ell-i-1)!\Big]^2}, \quad  i= 0,\ldots, \ell-3
$$
$$
     \tau_{ij+2}= \frac{  4(-1)^{2\ell-i-j-1}\,(\ell-i)\, (j+2-\ell)(2\ell-1)!
     }{\,(i)! \,(j+2)!\, (2\ell-i)!\, (2\ell-j-2)!}, \quad  i= 0,\ldots,
     \ell-3,\,\,\,\, j=i,\ldots, \ell-3,
     $$
     $$
  \lambda_{ij}=- \omega_{ij}=
  \frac{  2(-1)^{2\ell-i-j-1}(2\ell-1)! }{(i)!(j+1)!(2\ell-i-1)!(2\ell-j-2)!},
  \,\   i= 0,\ldots, \ell-2, \, j= i,\ldots, \ell-2,
  $$
  $$
      \varsigma_{ij}= \frac{  -4(-1)^{2\ell+i-j-1}\,(\ell-i)
      (\ell-j-1)\,(2\ell-1)! }{(i)!\,(j+1)!\,(2\ell-i)!\,(2\ell-j-1)!},\,\ i=
      0,\ldots, \ell-2,\,\, j= i,\ldots, \ell-2,
      $$
      $$
      \varepsilon_{ij}= \frac{  2(-1)^{\ell-i-j-2}\,\,(2\ell-1)!
      }{(i)!\,(j+2)!\,(2\ell-i-1)!\,(2\ell-j-3)!},\quad i= 0,\ldots,
      \ell-3,\,\, j= i,\ldots, \ell-3,\\ 
  $$
\end{thm}

\section{Conclusion}

In this short paper, we have constructed, via the search algorithm presented in
\cite{Als17}, the Casimir operators of certain
conformal Galilei algebras with central extensions. The focus was on the cases
corresponding to underlying spatial dimensions $d=1$
and $d=2$, and different values of $\ell$ (positive half-odd integer or integer).
The modest goal of the article was to draw attention to the fact that the
algorithm can be augmented by making use of a structural feature of these
algebras, namely the algebra anti-automorphism $\omega$, that allowed us to
restrict the search algorithm even further to only look for Casimir operators
conjugate under the action of $\omega$. 

It would be interesting to look at further cases corresponding to higher
spatial dimension $d \geq 2 $ for arbitrary values of $\ell$. More generally,
kinematical Lie algebras, among them the conformal and Schrodinger algebras,
are still object of an ongoing classification (see in particular the recent
works of Figueroa-O'Farrill and Andrzejewski \cite{Fig18e1,AndFig18,Fig18e2}). The method
developed in the current paper may enable further insight into the structure of
Casimir operators of a wider class of non-semisimple Lie algebras.

{\bf Acknowledgements:}

The research of FA is supported by Prince Sattam Bin Abdulaziz University. IM is supported
by the Australian Research Council Discovery Grant DP160101376. PSI is supported by the Australian Research Council
through Discovery Project DP150101294.

\end{document}